\documentclass[12pt,twoside]{article}
\usepackage{amsmath}

\newcommand{\ts}{\thinspace}
\newcommand{\dd}{\mathop{\mathrm d\null}\nolimits\!}

\input BoxedEPS
\SetRokickiEPSFSpecial 
\HideDisplacementBoxes 

\begin{document}
\pagestyle{myheadings}
\markboth{R. Jackiw}{What Good Are 
Quantum Field Theory Infinities?}

 \oddsidemargin=0.75in
 \evensidemargin=0.25in

\title{What Good Are\\
Quantum Field Theory Infinities?}

 \author{Roman Jackiw\thanks{This work is supported in part by funds
provided by  the U.S.~Department of Energy (D.O.E.) under contract
\#DE-FC02-94ER40818.\quad MIT CTP\# 2918\quad November 1999\break}\\
\small \it Center for Theoretical Physics\\
\small \it Massachusetts Institute of Technology\\
\small \it Cambridge, MA ~02139--4307\\[2ex]
\small Mathematical Physics 2000\\
\small Imperial College Press, London, UK}

\date{}

\maketitle

\vspace*{-0.25in}

{\narrower
\abstract{\noindent\small
A lesson for the new millennium from quantum field theory: Not
all field-theoretic infinities are bad. Some give rise to finite, 
symmetry-breaking effects,  whose consequences are observed in
Nature.\par}\par}
\vspace{0.25in}

\noindent
\noindent
Quantum field theory is the most successful theoretical structure in
physics, with applications that range from the short distances of subatomic particles to
the microscopic dimensions characterizing atomic, chemical, and condensed
matter physics, and onto the astronomical distances where quantum field
theory fuels ``inflation'' -- a speculative but completely physical analysis of
early universe cosmology.  Remarkably, no experimental observation has
contradicted the predictions that are made by appropriate field theoretical
models for the relevant phenomena.   When accurate calculation is feasible
and precision experiments are available, numerical agreement between theory
and experiment extends to many significant places, as for example in the
ground-state energies of simple atoms like hydrogen and helium, or in the
magnetic moments of electrons and muons.

Nevertheless, this gloriously successful invention of the human mind  is
logically defective in that some well-posed questions cannot be answered -- a
computation that should resolve the question can yield ambiguous or
meaningless answers.  This happens because the available methods of
calculation encounter infinities that either persist, leading to meaningless
results, or cancel among themselves, leaving ambiguous, undetermined
``finite'' parts.  There is no reason to suppose that this defect should be
attributed to the method of (approximate) computation -- it appears to be
intrinsic to interacting quantum field theory when excitations are point
particles and interactions are local.  (More specifically, I am referring to
ultraviolet infinities, which arise because various integrals over
intermediate energies diverge at their high-energy end, which corresponds to
short distances in position space.  There are also other infinities, like
diverging perturbative series or infrared/large-distance singularities
associated with long-range forces.  But these infinities are less
troublesome, because they are attributed to the approximation method and are
not viewed as intrinsic defects of quantum field theory.)

For physically relevant models, but not including gravity theory, it has been
possible to isolate the infinities by the ``renormalization'' procedure, which
hides them and also permits unambiguous calculation of quantities not
contaminated by the infinities.  Within this framework  definite
numerical results have been obtained, which in principle explain all observed
fundamental processes.  (Failure to tame infinities in quantum gravity has
thus far been irrelevant for practical purposes, because all presently
observed manifestations of gravitational forces are described by the classical
Newton-Einstein theory.)

In spite of the great success of quantum field theory, its
infinities notwithstanding, there are many who remain unconvinced by the
pragmatism of renormalization.  Dirac and Schwinger, who count among the
creators of quantum field theory and renormalization theory, respectively,
ultimately rejected their constructs because of the infinities.
But even  those who accept renormalization  disagree  about
its ultimate efficacy at well-defining a theory.  Some argue that sense can be
made only of ``asymptotically free'' renormalizable field theories -- in these
theories the interaction strength decreases with increasing energy.  On the
contrary, it is claimed that asymptotically nonfree models, like
electrodynamics and
$\phi^4$-theory, do not define quantum theories, even though they are
renormalizable -- it is said ``they do not exist.''  Yet electrodynamics
is the most precisely verified quantum field theory, while the $\phi^4$-model
is a necessary component of the ``standard model'' for elementary particle
interactions, which thus far has met no experimental contradiction.

The ultraviolet infinities appear as a consequence of space-time localization
of excitations and of their interactions.  (Sometimes it is claimed that field-theoretic 
infinities arise from the unhappy union of quantum theory with special relativity. 
But this does not describe all cases -- later I shall discuss a nonrelativistic,
ultraviolet-divergent, and renormalizable field theory.)  Therefore choosing models
with extended excitations and  interactions provides a way for avoiding ultraviolet
infinities.  These days ``string theory'' is a model with precisely such extended
features, and all quantum effects -- including gravitational ones -- are ultraviolet
finite.  This very desirable state of affairs has persuaded many that fundamental
physical theory in the next millennium should be based on the string paradigm
(generalized to encompass even more extended structures, like membranes and so
on).  This will replace quantum field theory, which although marred by its ultraviolet
infinities has served us well in the twentieth century.

My goal in this essay is to argue that at least some of the divergences of
quantum field theory must not be viewed as unmitigated defects. On the
contrary, they convey crucially important information about the physical
situation, without which most of our theories would not be physically
acceptable.  The stage where my unconventional considerations play a role is
that of symmetry, symmetry breaking, and conserved quantum numbers, so next I
have to review these ideas.

Physicists are mostly agreed that ultimate laws of Nature enjoy a high degree
of symmetry.  Presence of symmetry implies
absence of complicated and irrelevant structure, and our conviction that this
is fundamentally true reflects an ancient aesthetic prejudice -- physicists
are happy in the belief that Nature in its fundamental workings is
essentially simple.  Moreover, there are practical consequences of the
simplicity entailed by symmetry -- it is easier to understand the predictions of
physical laws.  For example, working out the details of very-many-body motion
is beyond the reach of actual calculations, even with the help of
computers.  But taking into account the symmetries that are present allows
understanding at least some aspects of the motion, and charting regularities
within it.

Symmetries bring with them conservation laws -- an association that is
precisely formulated by Noether's theorem.  Thus time-translation symmetry,
which states that physical laws do not change as time passes, ensures energy
conservation; space-translation symmetry,  the statement that physical laws
take the same form at different spatial locations, ensures momentum
conservation.  For another example, we note that the quantal description makes
use of complex numbers.  But physical quantities are real,
so complex phases can be changed at will, without affecting physical content. 
This invariance against phase redefinition, called \emph{gauge symmetry}, leads
to charge conservation.  The above examples show that symmetries
are linked to constants of motion.  Identifying such constants on the one
hand satisfies our urge to find regularity and permanence in natural
phenomena, and on the other hand we are provided with useful markers for
ordering physical data.

Moreover, a large degree of symmetry in the mathematical formulation of 
physically successful quantum field theory models is desireable not
only  aesthetically but also  practically.   Symmetry
facilitates unraveling the consequences of the complicated dynamical model;
more importantly, the presence of symmetry is required for a successful
renormalization of the infinities, so that unambiguous answers can be
extracted from the formalism.

However, in spite of our preference that descriptions of Nature be enhanced
by a large amount of symmetry and characterized by many conservation laws, we
must   recognize  that actual physical phenomena rarely exhibit overwhelming
regularity.  Therefore, at the very same time that we construct a physical
theory with intrinsic symmetry, we must find a way to break the symmetry in
physical consequences of the model.  Progress in physics can be frequently
seen as the resolution of this tension.

In classical physics, the principal mechanism for symmetry breaking, realized
already within Newtonian mechanics, is through boundary and initial
conditions on dynamical equations of motion. For example, radially
symmetric dynamics for planetary motion allows radially nonsymmetric,
noncircular orbits with appropriate initial conditions.  But this mode of
symmetry breaking still permits symmetric configurations -- circular orbits,
which are rotationally symmetric, are allowed.  In quantum mechanics, which
anyway does not need initial conditions to make physical predictions, we must
find mechanisms that prohibit symmetric configurations altogether.

In the simplest, most direct approach to symmetry breaking, we suppose that in
fact dynamical laws are not symmetric, but that the asymmetric effects are
``small" and can be ignored ``in first approximation."  Familiar examples  are the
breaking of rotational symmetry in atoms by an external electromagnetic field or of
isospin symmetry by the small electromagnetic interaction. 
However, this explicit breaking of symmetry is without fundamental interest
for the exact and complete theory; we need more intrinsic mechanisms that work
for theories that actually are symmetric.

A more subtle idea is \emph{spontaneous symmetry breaking}, where the dynamical
laws are symmetric, but only asymmetric configurations are actually realized (because
the symmetric ones are energetically unstable).  This mechanism, urged on particle
physicists by Heisenberg, Anderson, Nambu, and Goldstone, is readily illustrated by
the potential energy profile possessing left-right symmetry and depicted in the
Figure.  The left-right symmetric value at the origin is a point of unstable
equilibrium; stable equilibrium is attained at one of the two reflection-unsymmetric
points
$\pm a$. Moreover, in quantum field theory, the energy barrier separating the two
asymmetric configurations is infinite and no tunneling occurs between them. Once the
system settles in one or the other location, left-right parity is absent.  One says that
the symmetry of the equations of motion is ``spontaneously'' broken by the stable
solution.
\bigskip

\centerline{Energy density}
\bigskip
\centerline{\BoxedEPSF{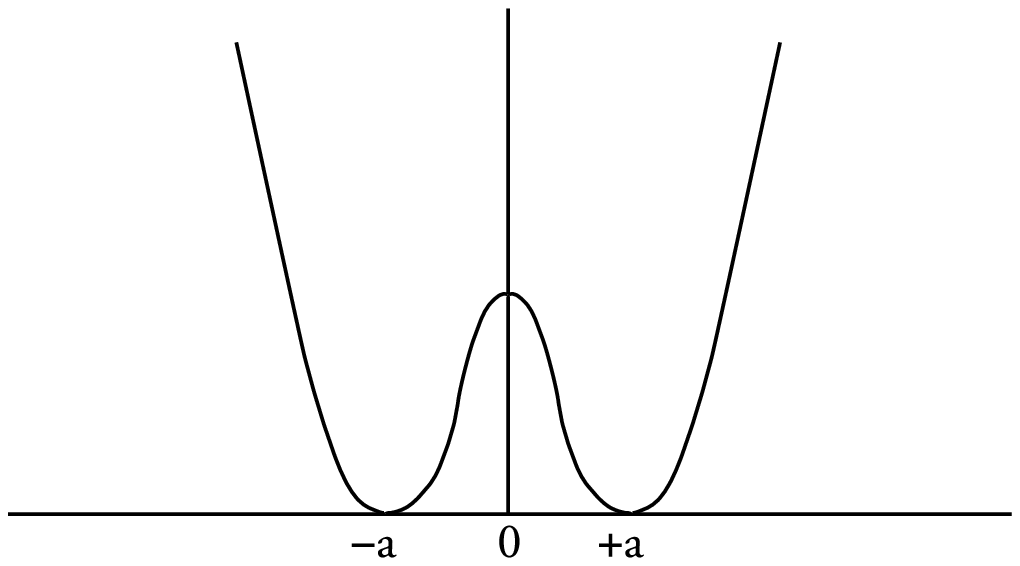 scaled 1000}}
\bigskip

{\small\narrower\noindent Left-right symmetric energy
density.  The symmetric point at 0 is energetically unstable.  Stable
configurations are at $\pm a$.  Because field theory is defined in an
infinite volume, the finite energy density separating $\pm a$ produces an
infinite energy barrier and tunneling is suppressed.  The system settles into
state $+ a$ or $- a$ and left-right symmetry is spontaneously broken.\par}
\bigskip

While the predictions of a theory with spontaneously broken symmetry no longer
follow the patterns that one would find if the symmetry were present in the
solutions, one important benefit of the symmetry remains: the renormalization
procedure is unaffected.  So the mechanism of
spontaneous symmetry breaking accomplishes the phenomenologically desired
reduction of formal symmetries without endangering renormalization, but it
does not reduce them enough.  Fortunately there exists a further, even more
subtle mode of symmetry breaking, with which we can further suppress
symmetries, thereby bringing our theories in accord with observed phenomena. 
Here one crucially relies on the various ultraviolet infinities of local
quantum field theory, for which the renormalization procedure (needed
to make sense of the theory) cannot be carried out in a manner consistent
with the symmetry.  Nevertheless the symmetry breaking effects are finite,
even though they arise from infinities.

This mode of symmetry breaking is called \emph{anomalous} or \emph{quantum
mechanical}, and in order to explain it, let me begin by recalling that the
quantum revolution did not erase our reliance on the earlier, classical
physics.  Indeed, when proposing a theory, we begin with classical concepts
and construct models according to the rules of classical, prequantum
physics.  We know, however, such classical reasoning is not in accord with
quantum reality.  Therefore, the classical model is reanalyzed by the rules
of quantum physics (which comprise the true laws of Nature), that is, the
classical model is quantized.

Differences between the physical pictures drawn by a classical description and
a quantum description are of course profound.  To mention the most dramatic,
we recall that dynamical quantities are described in quantum mechanics by
operators, which need not commute.  Nevertheless, one expects that some
universal concepts transcend the classical/quantal dichotomy, and enjoy
rather the same role in quantum physics as in classical physics.

For a long time it was believed that symmetries and conservation laws of a
theory are not affected by the transition from classical to quantum rules. 
For example, if a model possesses translation and gauge invariance on the
classical level, and consequently energy/momentum and charge are conserved
classically, it was believed that after quantization the quantum model is
still translation and gauge invariant so that the energy/momentum and charge
operators are conserved within quantum mechanics, that is, they commute with
the quantum Hamiltonian operator.  But now we know that in general this need
not be so.  Upon quantization, some symmetries of classical physics may
disappear when the quantum theory is properly defined in the presence of its
infinities.  Such tenuous symmetries are said to be \emph{anomalously} broken; 
although present classically, they are absent from the quantum version of the
theory, unless the model is carefully arranged to avoid this effect.

The nomenclature is misleading.  At its discovery, the phenomenon was
unexpected and dubbed ``anomalous."  By now the surprise has worn off, and the
better name today is ``quantum mechanical" symmetry breaking.

Anomalously or quantum mechanically broken symmetries play several and
crucial roles in our present-day physical theories.  In
some instances they save a model from possessing too much symmetry,
which would not be in accord with experiment.  In other instances the
desire to preserve a symmetry in the quantum theory places strong constraints
on model building and gives experimentally verifiable predictions; more about
this later.\cite{ref:1}

Now I shall describe two specific examples of the anomaly phenomenon. 
Consider first massless fermions moving in the 
background of an electromagnetic field.  Massive,
spin-$1\over 2$ fermions possess two spin states -- up and down -- but massless
fermions can exist with only one spin state (out of two), called a \emph{helicity}
state, in which spin is projected along (or against) the direction of motion.  So the
massless fermions with which we are here concerned carry only one helicity and
these are an ingredient in present-day theories of quarks and leptons.  Moreover,
they also arise in condensed matter physics, not because one is dealing with massless,
single-helicity particles, but because a well-formulated approximation to
various many-body Hamiltonians can result in a first-order matrix equation
that is identical to the equation for single-helicity massless fermions, that is, a
massless Dirac-Weyl equation for a spinor $\Psi$.

If we view the spinor field $\Psi$ as an ordinary mathematical function, we
recognize that it possesses a complex phase, which can be altered without
changing the physical content of the equation that $\Psi$ obeys.  We expect
therefore that this instance of gauge invariance implies charge conservation. 
However, in a quantum field theory $\Psi$ is a quantized field operator, and
one finds that in fact the charge operator $Q$ is not conserved; rather
$$
{\dd Q \over \dd t} = {i \over \hbar} [H,Q] \propto \int_{\mathrm{volume}}
\hbox{\bf E} \cdot \hbox{\bf B}
$$  
where {\bf E} and {\bf B} are the background electric and magnetic fields in
which our massless fermion is moving -- gauge invariance is lost! 

One way to understand this breaking of symmetry is to observe that our model
deals with \emph{massless} fermions and conservation of charge for 
\emph{single-helicity} fermions makes sense only if there are no fermion masses. 
But quantum field theory is beset by its ultraviolet infinities, which must be
controlled in order to do a computation.  This is accomplished by regularization and
renormalization, which introduces mass scales for the fermions, and we see that the
symmetry is anomalously broken by the ultraviolet infinities of the theory.

The phase-invariance of single-helicity fermions is called \emph{chiral (gauge)
symmetry}, and chiral symmetry has many important roles in the standard model,
which involves many kinds of fermion fields, corresponding to the various
quarks and leptons.  In those channels where a gauge vector meson couples to
the fermions, chiral symmetry must be maintained to ensure gauge invariance. 
Consequently, fermion content must be carefully adjusted so that the anomaly
disappears.  This is achieved because the proportionality constant in
the above failed conservation law involves a sum over all the
fermion charges, $\sum\limits_n q_n$, so if that quantity vanishes the
anomaly is absent.  In the standard model the sum indeed vanishes, separately
for each of the three fermion families.  For a single family this works out
as follows:
\bigbreak

\begin{center}
\begin{tabular}{lrlr}
three quarks & $  q_n$   =  ${2 \over 3}$ & $\Rightarrow$ & $2$ \\[1ex]
three quarks &  $  q_n$  =  $-{1 \over 3}$ & $ \Rightarrow$ & $ -1$ \\[1ex]
one~charged~lepton & $  q_n$ =  $-1$ & $ \Rightarrow $ & $-1$ \\[1ex]
one~neutrino~lepton & $  q_n$ =  $0 $ & $\Rightarrow$ & $ 0$ \\
\cline{2-4}
              &           $\sum\limits_n^{\phantom{0}} q_n$ &  =   & $ 0$
\end{tabular}
\end{center}
\bigbreak

In channels to which no gauge vector meson couples, there is no requirement
that the anomaly vanish, and this is fortunate. A theoretical analysis
shows that chiral gauge invariance in the up-down quark channel prohibits the
two-photon decay of the neutral pion (which is composed of up and down
quarks).  But the decay does occur with the
invariant decay amplitude of $0.025 \pm 0.001$\ts GeV${}^{-1}$. Before anomalous
symmetry breaking was understood, this decay could not be fitted into the
standard model, which seemed to possess the decay-forbidding
chiral symmetry.  Once it was realized that the relevant chiral symmetry is
anomalously broken, this obstacle to phenomenological viability of the standard
model was removed.  Indeed since the anomaly is completely known, the decay
amplitude can be completely calculated (in the approximation that the pion is
massless) and one finds
$0.025$\ts GeV${}^{-1}$, in excellent agreement with experiment.  

We must conclude that Nature knows about and makes use of the anomaly
mechanism. On the one hand fermions are arranged into gauge-anomaly--free
representations, and the requirement that anomalies disappear ``explains" the
charges of elementary fermions. On the other hand the pion decays into two
photons because of an anomaly in an ungauged channel.  It is therefore
paradoxical but true that in local quantum field theory these
phenomenologically desirable results are facilitated by ultraviolet
divergences, which give rise to finite symmetry anomalies, derived from
infinities.

The observation that infinities of quantum field theory lead to anomalous
symmetry breaking allows comprehending a second
example of quantum-mechanical breaking of yet another symmetry --  scale
invariance.  Like the space-time translations mentioned earlier, which lead to
energy-momentum conservation, scale transformations also act on space-time
coordinates, but in a different manner. They dilate the coordinates, thereby
changing the units of space and time measurements.  Such transformations will
be symmetry operations in models that possess no fundamental parameters with
time or space dimensionality, and therefore do not contain an absolute scale
for units of space and time.  Our quantum chromodynamical (QCD) model for
quarks is free of such dimensional parameters, and it would appear that this
theory is scale invariant -- but Nature certainly is not!  The observed
variety of different objects with different sizes and masses exhibits many
different and inequivalent scales.  Thus if scale symmetry of the
\emph{classical} field theory, which underlies the \emph{quantum} field theory
of QCD, were to survive quantization, experiment would have grossly
contradicted the model, which therefore would have to be rejected. 
Fortunately, scale symmetry is quantum mechanically broken, owing to the
scales that are introduced in the regularization and renormalization of
ultraviolet singularities.  Once again a quantum field-theoretic pathology has a
physical effect, a beneficial one --  an unwanted symmetry is anomalously
broken, and removed from the theory.

A different perspective on the anomaly phenomenon comes from
the path integral formulation of quantum theory, where one integrates over
classical paths the phase exponential of the classical action:
$$
{\rm Quantum~Mechanics} \Longleftrightarrow 
\int_{\rm(measure~on~paths)} e^{i {\rm
(classical~action)}/\hbar}\ .
$$
When the classical action possess a symmetry, the quantum theory will respect
that symmetry if the measure on paths is unchanged by the relevant
transformation.  In the known examples (chiral symmetry, scale symmetry)
anomalies arise precisely because the measure fails to be invariant and this
failure is once again related to infinities. The measure is an \emph{infinite}
product of measure elements for each point in the space-time where the quantum
(field) theory is defined; regulating this infinite product destroys its
apparent invariance.

Yet another approach to chiral anomalies, which arise in (massless)
fermion theories, makes reference to the first instance of
regularization/renormalization, used by Dirac to remove the negative-energy
solutions to his equation.  Recall that to define a quantum \emph{field}
theory of fermions, it is necessary to fill the negative-energy sea and to
renormalize the infinite mass and charge of the filled states to zero.  In
modern formulations this is achieved by ``normal ordering", but for our
purposes it is better to remain with the more explicit procedure of
subtracting the infinities, that is, renormalizing them.

It can then be shown that in the presence of an external gauge field, the
distinction between ``empty" positive-energy states and ``filled"
negative-energy states cannot be drawn in a gauge-invariant manner, for
massless, single-helicity fermions.  Within this framework, the chiral
anomaly comes from the gauge noninvariance of the infinite negative-energy
sea.  Since anomalies have physical consequences, we must assign physical
reality to this infinite negative-energy sea. 

Actually, in condensed matter physics, where a Dirac-type equation governs
electrons, owing to a linearization of dynamical equations near the Fermi
surface, the negative-energy states \emph{do} have physical reality. They
correspond to filled, bound states, while the positive energy states
describe electrons in the conduction band.  Consequently, chiral anomalies
also have a role in condensed matter physics, when the system is idealized so
that the negative-energy sea is taken to be infinite.  
 
In this condensed matter context another curious, physically realized, and
infinity-driven phenomenon has been identified.  When the charge of the filled
negative states is renormalized to zero, one is subtracting an infinite
quantity, and rules have to be agreed upon so that no ambiguities arise when
infinite quantities are manipulated.  With this agreed-upon subtraction
procedure, the charge of the vacuum is zero, and filled states of positive
energy carry integer units of charge.  Into the system one can insert a
soliton -- a localized structure that distinguishes between different domains
of the condensed matter.  In the presence of such a soliton, one needs to
recalculate charges using the agreed-upon rules for handling infinities and
one finds, surprisingly, a noninteger result, typically half-integer: the
negative-energy sea is distorted by the soliton to yield a half-unit of
charge.  The existence of fractionally charged states in the presence of
solitons has been experimentally identified in polyacetylene.  We thus have
another example of a physical effect emerging from infinities of quantum
field theory.\cite{ref:2}

Let me conclude my qualitative discussion of anomalies with an
explicit example from quantum mechanics, whose wave functions provide a link
between particle and field-theoretic dynamics.  My example also dispels any
suspicion that ultraviolet divergences and the consequent anomalies are tied to
the complexities of \emph{relativistic} quantum field theory. The
nonrelativistic example shows that locality is what matters.

Recall first the basic dynamical equation of quantum mechanics: the
time independent Schr\"odinger equation for a particle of mass $m$ moving
in a potential $V$({\bf r}) with energy $E$:
$$
\Bigl( - \nabla^2 + \frac{2m}{\hbar^2} V ({\bf r}) \Bigr) \psi
({\bf r}) =  \frac{2m}{\hbar^2} E \psi ({\bf r})  \  .
$$
In its most important physical applications, this equation is taken in three
spatial dimensions and $V ({\bf r})$ is proportional to $1/r$ for the
Coulomb force relevant in atoms.  Here we want to take a different model with
potential that is proportional to the inverse square, so that the
Schr\"odinger equation is presented as
$$
\Bigl( - \nabla^2 + {\lambda \over r^2} \Bigr) \psi({\bf r})  = 
k^2  \psi (\hbox{\bf r})  \ , \quad 
k^2 \equiv {2m \over \hbar^2}E \,\, .
$$
In this model, transforming the length scale is a symmetry: because the
Laplacian scales as $r^{-2}$, $\lambda$ is dimensionless and in
the above there is no intrinsic unit of length.  A consequence of scale
invariance is that the scattering phase shifts and the $S$ matrix, which in
general depend on energy, that is, on $k$, are energy independent in scale-invariant
models. And indeed when the above Schr\"odinger equation is solved, one verifies
this prediction of the symmetry by finding an energy-independent
$S$ matrix.  Thus scale invariance is maintained in this example -- there are
no surprises.

Let us now look to a similar model, but in two dimensions with a
$\delta$-function potential, which localizes the interaction at a point:
$$
\Bigl( - \nabla^2 + \lambda \delta^2 ({\bf r}) \Bigr) \psi
(\hbox{\bf r}) =  k^2\psi ({\bf r})  \  .
$$
Since in two dimensions the two-dimensional $\delta$-function scales as
$1/r^2$, the above model also appears scale invariant; $\lambda$ is
dimensionless.  But in spite of the simplicity of the local contact
interaction, the Schr\"odinger equation suffers a short-distance, ultraviolet
singularity at {\bf r}=0, which must be renormalized.  Here is not the place
for a detailed analysis, but the result is that only the $s$-wave possesses a
nonvanishing phase shift $\delta_0$, which shows a logarithmic dependence on
energy:
$$
\cot \delta_0 = {2 \over \pi} \ln kR + {1 \over \lambda}
$$
$R$ is a scale that arises in the renormalization, and scale symmetry is
decisively and quantum mechanically broken.   The scattering is nontrivial
solely as a consequence of anomalously broken scale invariance. (It is easily verified
that the two-dimensional $\delta$-function in classical theory, where  there
are no anomalies and it \emph{is} scale invariant, produces no scattering.) 
To make sense of the above phase shift in the limit
$R\to \infty$, one must ``renormalize'' the bare coupling constant $\lambda$,
allowing it to depend on $R$ in just such a way that $\cot \delta_0$ is
$R$-independent (for large~$R$).  Alternatively, one recognizes that the $S$
matrix $e^{2i\delta_0}$ possesses a pole, corresponding to a bound state with
energy
$$
E_B = -\frac{\hbar^2}{2mR^2} e^{-\pi/\lambda}\ .
$$
Therefore one may reexpress $\delta_0$ in terms of $E_B$, rather than $R$. 
With this substitution, dependence on $\lambda$ disappears (as it must since
$\lambda$ is $R$-dependent) and the dimensionless (infinite) coupling
constant $\lambda$, has been traded for a dimensional and physical parameter~$E_B$:
$$
\cot \delta_0 = \frac{1}{\pi} \ln \frac{E}{ |E_B|}
$$

Similar anomalous breaking of scale invariance occurs in relativistic field
theory, and perhaps explains the appearance of a dimensional mass parameter
in QCD as a replacement for the dimensionless, but renormalization dependent,
coupling constant.\cite{ref:3}

I believe that as the millennium draws to a close, and we look forward eagerly 
to the new physics ideas that will flourish in the new era, one very important
lesson we should take from quantum field theory is not to
banish all its infinities.  Apparently the
mathematical language with which we are describing Nature cannot account for
all natural phenomena in a clear fashion. Recourse must be made to
contradictory formulations involving \emph{infinities}, which nevertheless
lead to accurate descriptions of experimental facts in \emph{finite} terms. 
It will be most interesting to see how string theory and its evolutions,
which purportedly are completely finite and consistent, will handle this
issue, which has been successfully, if paradoxically, resolved in quantum field
theory.

\end{document}